\documentclass[12pt]{iopart}
\usepackage{graphicx} 
\usepackage{psfrag}
\begin{document}
% Journal identifier can be put here if required, e.g.
\jl{1}

\title{Dimensional crossover in dipolar magnetic layers}

\author{M Bulenda \dag , U C T\"auber \ddag\ and F Schwabl \dag 
\footnote[3]{To
whom correspondence should be addressed.}}

\address{\dag \ Institut f\"ur Theoretische Physik, Physik--Department der
Technischen Universit\"at M\"unchen,  James--Franck--Stra\ss e, 
D--85747 Garching, Germany}
\address{\ddag\ Physics Department, Virginia Polytechnic Institute and State
University, Blacksburg, VA 24061--0435, USA}

\begin{abstract}
We investigate the static critical behaviour of a uniaxial magnetic layer,
with finite thickness $L$ in one direction, yet infinitely extended in the 
remaining $d$ dimensions.
The magnetic dipole-dipole interaction is taken into account. 
We apply a variant of Wilson's momentum shell renormalisation group approach to
describe the crossover between the critical behaviour of the 3-D Ising, 2-d 
Ising, 3-D uniaxial dipolar, and the 2-d uniaxial dipolar universality classes.
The corresponding renormalisation group fixed points are in addition to 
different effective dimensionalities characterised by distinct analytic 
structures of the propagator, and are consequently associated with varying 
upper critical dimensions.  
While the limiting cases can be discussed by means of dimensional $\epsilon$ 
expansions with respect to the appropriate upper critical dimensions, 
respectively, the crossover features must be addressed in terms of the 
renormalisation group flow trajectories at fixed dimensionality $d$.
\end{abstract}


%\submitted

%\maketitle

\section{Introduction}

Layered magnetic systems are of obvious technological importance, e.g., in
magnetic storage devices.
Presently, magneto-optic storage devices based on amorphous GdTbFe and TbFeCo 
alloys are in commercial use, allowing for storage densities of up to 
$10^8$ bits/cm$^2$. In the future, however, Pt/Co, Pd/Co, and Co/Au multilayers
of a few Angstrom thickness are expected to lead to four times as high storage 
capacities.
On the purely theoretical side, they furthermore represent an intriguing model 
system with the competing exchange- and dipolar-dominated universality classes 
on the one hand, and the dimensional crossover between three and two effective
dimensions on the other hand.
The mathematical description of the associated crossover features within the
framework of the renormalisation group theory is technically challenging, as it
involves both varying analytic structures of the propagators, and different 
upper critical dimensions for the associated renormalisation group fixed 
points.
The aim of the present paper is to demonstrate that such a unifying crossover
description is feasible, even with the aforementioned difficulties at hand.
For a layered uniaxial magnetic system of finite width $L$ in one direction
and with  
dipolar interactions taken into account, the renormalisation program 
is explicitly performed to one-loop order for the entire crossover regime by 
means of a variant of Wilson's momentum shell method at fixed transverse 
dimension $d$.
Moreover, it is shown that the scheme described here encompasses all the
relevant universality classes involved, and the critical exponents at each
renormalisation group fixed point may be computed within an $\epsilon$ 
expansion about the respective upper critical dimensions.

In the theory of phase transitions there has been long-standing interest in the
critical properties of finite systems as well as in the influence of the 
magnetic dipole-dipole interaction on static and dynamic critical behaviour. 
The first investigations for the crossover behaviour between the different 
critical regimes in dipolar systems date back thirty years:
In their pioneering work, Larkin and Khmel'nitski\u{i} \cite{Larkin} showed 
that the upper critical dimension of a uniaxial magnet with magnetic 
dipole-dipole interaction is reduced by one as compared to the pure Ising 
system. 
The mathematical description of the crossover in a three-dimensional system 
with easy axis and dipolar forces therefore necessitated a renormalisation 
scheme capable of dealing with varying upper critical dimensions associated 
with the different non-trivial fixed points.
Modifying the generalised dimensional regularisation method by Amit and 
Goldschmidt \cite{Amit2}, such a description was developed in Ref.~\cite{Frey1}
for the crossover between the Ising and 3-d uniaxial dipolar fixed points of a 
uniaxial dipolar magnet, and in Ref.~\cite{Frey4} for the crossover from 
isotropic to directed percolation. 
This approach was then applied to an even more complex scenario by Ried et al.
\cite{Ried1}, who considered the crossovers between all four possible fixed 
points of a uniaxial dipolar system, taking the finiteness of the anisotropy 
into account. 
It should be noted, though, that all these calculations employed the harmonic
dipolar propagator originally derived in Ref.~\cite{Aharony2}. 
Yet this propagator is adequate only in the case of a three-dimensional 
infinite system, or, more generally, for any system of the same dimensionality 
as the one used for computing the effective magnetic dipole-dipole interaction.
It is therefore {\it not} applicable for 3-d electrodynamics, if the system 
itself is two-dimensional, for example, or is finite in one or more directions.

Static critical behaviour in finite systems with purely short-range 
interactions was investigated in Refs.~\cite{Rudnick2,Brezin,Esser1,Chen1}. 
Here, the asymptotic critical behaviour was addressed, and no crossover scaling
functions were calculated. 
Dimensional crossovers were studied in the context of quantum phase transitions
by Chakravarty, Halperin, and Nelson for quantum antiferromagnets 
\cite{Chakravarty1}, and by O'Connor and Stephens for an Ising system between 
four and three dimensions \cite{OConnor1}.
Yet we are not aware of any work concerning the critical behaviour in a finite 
system that takes into account the magnetic dipole-dipole interaction.

In this paper we go beyond the cited literature in describing the crossover
scenario for a uniaxial magnetic system with dipolar interactions that is
infinitely extended in $d$ dimensions, but has a finite thickness $L$ in one 
additional direction. 
For simplicity, we assume that the anisotropy energy is larger than all other 
relevant energy scales, and hence only take into account the Ising-type up-down
symmetry. 
In the case of $d=2$, such a system is supposed to exhibit four different 
regimes of non-trivial critical behaviour, in addition to the classical one
described by mean-field critical exponents; these are described by the 3-D
Ising, 2-d Ising, 3-D uniaxial dipolar, and 2-d uniaxial dipolar universality 
classes, respectively.
As will be shown, these regimes are characterised in addition to varying
effective system dimensions by different analytic forms of the propagator,
which originate in the dipole-dipole interaction.
Consequently, three different upper critical dimensions $d_c$ emerge: 
For the Ising systems, $D_c=4$ and $d_c=4$, respectively, for the 3-D uniaxial 
dipolar system $D_c=3$, and finally the upper critical dimension for the truly 
asymptotic critical behaviour of a 2-d uniaxial dipolar system with an easy 
axis in the system's plane will be demonstrated to be $d_c=\frac{7}{2}$. 

This paper is organised as follows: In section \ref{Propagator} we introduce 
the effective hamiltonian (free energy functional) and specifically the 
propagator (two-point correlation function in Gaussian approximation) that is 
to be used for our calculations.
In section \ref{method} we explain our variant of the momentum shell 
renormalisation method devised to mathematically describe the crossover 
scenarios.
In section \ref{ren} the resulting renormalisation group flow equations are 
derived to one-loop order and discussed.
Section \ref{Limit} contains the resulting limiting cases within the $\epsilon$
expansions in the vicinities of the appropriate upper critical dimensions. 
In section \ref{crossover} the crossovers between the critical fixed points are
discussed in terms of the renormalisation group trajectories. 
We conclude with a summary and brief discussion in section \ref{summary}.

\section{Effective hamiltonian}
\label{Propagator}

At first we need to deduce an effective free energy functional that is capable
of capturing the crossover features we are aiming at.
However, at this point we encounter the difficulty that the analytic forms of 
the propagators (Gaussian two-point correlation functions) describing uniaxial 
dipolar magnetic systems are quite different depending on the dimensionality 
$D = d+1$ of the system. 
We employ periodic boundary conditions to avoid surface effects, in which we 
are not specifically interested here. 
Rather we are concerned with the effect of the system's limited extent in one 
space direction as such.

Denoting the part of the wave vector ${\bi k}$ in the direction of the easy 
axis of the magnetisation as ${\bi q}$, and the perpendicular component as
${\bi p}$, ${\bi k} = {\bi q} + {\bi p}$, the propagators look as follows: 
In the case of a three-dimensional system, the relevant part of the inverse 
propagator is of the form $G_0({\bi q},{\bi p})^{-1} = r + c \, k^2 + g \, q^2
/ p^2$ \cite{Aharony1}, while for a two-dimensional system containing the easy 
axis, $G_0({\bi q},{\bi p})^{-1} = r + c \, k^2 + g \, q^2 / k$, with $k = 
|{\bi k}|$ \cite{Maleev}. 
On the other hand, if the easy axis is aligned perpendicular to the 2-d plane, 
the result is of the form $G_0({\bi q},{\bi p})^{-1} = r + c \, k^2 - g \, k$ 
\cite{Maleev}.
Here $r = T - T_c^0$ denotes the temperature distance to the mean-field 
transition temperature $T_c^0$, $c$ is a constant describing the stiffness of 
the system with respect to inhomogeneous order parameter configurations, and 
$g$ measures the strength of the dipole-dipole interaction as compared to the 
exchange interaction.

As our task is the description of a system with finite thickness, we have to
apply some care in the derivation of the appropriate form of the propagator. 
The contribution to the free energy stemming from the dipole-dipole interaction
between the magnetic moments connected to the spin variables $\bi S_{\bi R}$ at
lattice sites $\bi R$ reads 
\begin{eqnarray}
W_D = g\sum_{\bi R_1\ne\bi R_2}\sum_{\alpha\beta}
A^{\alpha\beta}(\bi R_1,\bi R_2) \ S_{\bi R_1}^\alpha S_{\bi R_2}^\beta \ ,
\end{eqnarray}
with the dipole tensor in real space being defined as
\begin{eqnarray}
A^{\alpha\beta}(\bi R_1,\bi R_2) & = & 
\frac{\delta^{\alpha\beta}}{|\bi R_1-\bi R_2|^3} - 
\frac{3(\bi R_1-\bi R_2)^\alpha(\bi R_1-\bi R_2)^\beta}{|\bi R_1-\bi R_2|^5}\ .
\end{eqnarray}
Here, the indices $\alpha$ and $\beta$ denote real space vector components.
In Fourier space, the dipole sum for a cubic lattice becomes
\begin{eqnarray}
A^{\alpha\beta}(\bi k) & = & A^{\alpha\beta}_{2d}(\bi p) + 
\frac{2\pi}{a_0} \sum_{\rho\ne 0} \rme^{iq\rho} \sum_{\bi g} 
\frac{(\bi p-\bi g)^\alpha (\bi p-\bi g)^\beta}{|\bi p-\bi g|} \ 
\rme^{-|\rho||\bi p-\bi g|} \ ,
\end{eqnarray}
if the spins are oriented in the plane, and
\begin{eqnarray}
A^{\alpha\beta}(\bi k)
& = & A^{\alpha\beta}_{2d}(\bi p) -
\frac{2\pi}{a_0} \sum_{\rho\ne 0} \rme^{iq\rho} \sum_{\bi g} |\bi p-\bi g| 
\, \rme^{-|\rho| \, |\bi p-\bi g|} \ ,
\end{eqnarray}
if the spins are oriented perpendicular to the plane \cite{Maleev}. 
$A^{\alpha\beta}_{2d}(\bi p)$ represents the dipole tensor for the 
two-dimensional system, with wave vectors ${\bi p}$ and associated 2-d 
reciprocal lattice vectors $\bi{g}$.
$\rho$ is the discrete coordinate of a lattice plane along the finite 
direction, and $a_0$ and $q$ indicate the corresponding 
lattice constant and wave numbers, respectively.
For the later treatment by means of the renormalisation group technique, we
need only retain the relevant long-wavelength contributions, and may thus
simplify the above expressions by restricting the reciprocal wave vector sums  
to $\bi{g}=0$ only.
This part yields the correct limits for the propagators of both the three- and 
the two-dimensional system.
 
Finally, we expand the resulting expressions to obtain the long-wavelength 
limits. 
For the easy axis lying in the plane we find the following form: 
\begin{eqnarray}
A^{\alpha \beta}(\bi{q},\bi{p}) & = & (1+f_1) \, \frac{p^\alpha p^\beta}{p} + 
f_2 \, \frac{p^\alpha p^\beta}{p} \, q^2 + f_3 \, p^\alpha p^\beta \nonumber \\
&&+ f_4 \, p^\alpha p^\beta \, p + f_5 \, p^\alpha p^\beta \, q^2 + 
{\cal O}(p^3 q^2, p^4, p^2q^4) \ .
\label{propent}
\end{eqnarray}
Denoting with $2N+1$ the number of lattice planes in the finite direction,
$L = 2 N \, a_0$, the $N$-dependent functions $f_i$ read:
\begin{eqnarray}
f_1&=&2N \ , \label{f1}\\
f_2&=&-\frac{a_0^2}{6} \, N\left(1+3N+2N^2\right) \ , \label{f2}\\
f_3&=&-a_0 \, N(1+N) \ , \label{f3}\\
f_4&=&\frac{a_0^2}{6} \, N\left(1+3N+2N^2\right) \ , \label{f4}\\
f_5&=&\frac{a_0^3}{4} \, N^2(1+N)^2 \ . \label{f5}
\end{eqnarray}
If the easy axis is aligned perpendicular to the plane, we arrive at
\begin{eqnarray}
A^{\alpha \beta}(\bi{q},\bi{p}) = - (1+{\tilde{f}}_1) \, p-{\tilde{f}}_3 \, p^2
- {\tilde{f}}_2 \, q^2 \, p + {\cal O}(p^3 q^2, p^4) \label{2.52}
\end{eqnarray}
instead, with the $N$-dependent functions
\begin{eqnarray}
{\tilde{f}}_1&=&2N \ , \\
{\tilde{f}}_2&=&-\frac{a_0^2}{6} \, N\left(1+3N+2N^2\right) \ , \\
{\tilde{f}}_3&=&-a_0 \, N(1+N) \ .
\end{eqnarray}
We note that the long-wavelength expansion does {\em not} commute with taking 
the limit of infinite system thickness, $N \to \infty$. 
Consequently, the above formulas are strictly valid only for finite system 
thickness $L$.
Nonetheless we shall see that the method to be described below will yield the
correct results even for the three-dimensional system. 

We are now in the position to write down the effective hamiltonians that serve
as starting points for our renormalisation group treatment. 
Dealing with a finite system in one direction, we symbolically write the sum
over all Fourier modes as
\begin{eqnarray}
\int_{\bi k} \ldots \equiv \frac{1}{(2 \pi)^d} \int_{0<|p|<\Lambda} d^d p \quad
\frac{1}{L} \sum_{q=\frac{2\pi}{L} n} \ldots \ , \quad n \in Z \ .
\label{int}
\end{eqnarray}
In the limit of the fully infinite system, $L \to \infty$, this definition is 
to be replaced with
\begin{eqnarray}
\int_{\bi k} \ldots \equiv \frac{1}{(2 \pi)^D} \int_{0<|p|<\Lambda} d^d p 
\int dq \ , \quad {\rm where} \quad D = d+1 
\label{Leonie}
\end{eqnarray}
denotes the dimension of the entire system.
In both situations, $\Lambda$ represents a cutoff whose numerical value is not 
relevant for the critical behaviour. 
Therefore we shall use $\Lambda=1$ throughout the paper (in appropriate 
dimensionless units). 
Notice that we only apply the cutoff for the wave vector components in the 
plane, and {\em not} for the perpendicular component. 

Thus the final effective free energy for a {\em layer with easy axis in the 
plane} assumes the form
\begin{eqnarray}
{\cal H}_{2d,\parallel} & = & \frac{1}{2} \int_{\bi{k}} \Big[ r_0 + c_0 \, k^2 
+ g (1+f_1) \, \frac{p_l^2}{p} + g f_2 \, \frac{p_l^2}{p} \, q^2 \nonumber \\
&&\qquad + g f_3 \, p_l^2 +g f_4 \, p_l^2 \, p + g f_5 \, p_l^2 \, q^2 \Big] \,
S_0(\bi{k}) S_0(-\bi{k}) \nonumber \\ 
&&+ \frac{u_0}{4!} \int_{\bi{k}_1}\int_{\bi{k}_2}\int_{\bi{k}_3}\int_{\bi{k}_4}
S_0(\bi{k}_1) S_0(\bi{k}_2) S_0(\bi{k}_3) S_0(\bi{k}_4) \
\delta\!\left(\sum_{i=1}^4 \bi{k}_i \right) \ ,
\label{Stompf}
\end{eqnarray}
where $q$ again is the wave vector component perpendicular to the system, and 
${\bi p}$ the in-plane part, with $p_l$ denoting the component pointing along 
the direction of the easy axis.
The hamiltonian for a {\em layer with easy axis perpendicular to the plane} 
reads
\begin{eqnarray}
{\cal H}_{2d,\perp} &=& \frac{1}{2} \int_{\bi{k}} \left[ r_0 + c_0 \, k^2 - g 
(1+{\tilde{f}}_1) \, p - g {\tilde{f}}_3 \, p^2 - g {\tilde{f}}_2 \, q^2 \, p 
\right] S_0(\bi{k}) S_0(-\bi{k}) \nonumber \\  
&&+ \frac{u_0}{4!} \int_{\bi{k}_1}\int_{\bi{k}_2}\int_{\bi{k}_3}\int_{\bi{k}_4}
S_0(\bi{k}_1) S_0(\bi{k}_2) S_0(\bi{k}_3) S_0(\bi{k}_4) \ 
\delta\!\left(\sum_{i=1}^4 \bi{k}_i \right) \ .
\label{Wicht}
\end{eqnarray}
For a {\em three-dimensional system} we find instead \cite{Aharony1}
\begin{eqnarray}
{\cal H}_{3d} &=& \frac{1}{2} \int_{\bi{k}} \left[ r_0 + c_0 \, k^2 + g \,
\frac{q^2}{p^2} \right] S_0(\bi{k}) S_0(-\bi{k}) \nonumber \\
&&+ \frac{u_0}{4!} \int_{\bi{k}_1}\int_{\bi{k}_2}\int_{\bi{k}_3}\int_{\bi{k}_4}
S_0(\bi{k}_1) S_0(\bi{k}_2) S_0(\bi{k}_3) S_0(\bi{k}_4) \ 
\delta\!\left(\sum_{i=1}^4 \bi{k}_i \right) \ ,
\label{3dHamilton}
\end{eqnarray}
where $q$ denotes the component of $\bi{k}$ along the direction of the easy 
axis, and $\bi{p}$ is the part of $\bi{k}$ perpendicular to the easy axis.

Thus, henceforth we are concerned with the description of critical phenomena 
with those three different hamiltonians. 
The free energy of the three-dimensional system cannot be recovered any more 
from the layered hamiltonian (\ref{Wicht}) by taking the limit of infinite 
layer thickness $N \to \infty$. 
Nevertheless we will show that all the limiting cases, including the 
three-dimensional uniaxial dipolar one, are described correctly within our
scheme.

\section{Renormalisation scheme}
\label{method}

As already mentioned, the expected critical regimes are clearly characterised 
by different system dimensions, as well as different analytical structures of 
the propagators, leading to different upper critical dimensions. 
Hence we need a renormalisation scheme capable of dealing with these different
fixed points in a unified manner, and sufficiently flexible as to describe the 
crossovers between them. 
In addition, we aim to capture the crossover features with the correct 
transition temperature $T_c(L,g)$ (see appendix A) of the finite system, and 
not with that of the infinite system as in 
Refs.~\cite{Rudnick2,Brezin,Esser1,Chen1}, where furthermore the studied 
systems were infinite in at most one direction (and consequently do not display
a proper phase transition). 
As our system asymptotically becomes two-dimensional, we have to circumvent 
problems with infrared divergences appearing in field-theoretic approaches, 
e.g., dimensional regularisation, for such low-dimensional systems
\cite{Amit2,Frey1,OConnor1,Amit1}.  
One more aspect is that we must include the full gradient terms in order to 
describe the crossover between Ising and uniaxial dipolar behaviour, and cannot
restrict ourselves to merely the most relevant parts in one of the limits as in
Ref.~\cite{Folk}.

\begin{figure}[t]
\begin{center}
\includegraphics[width=0.4\textwidth]{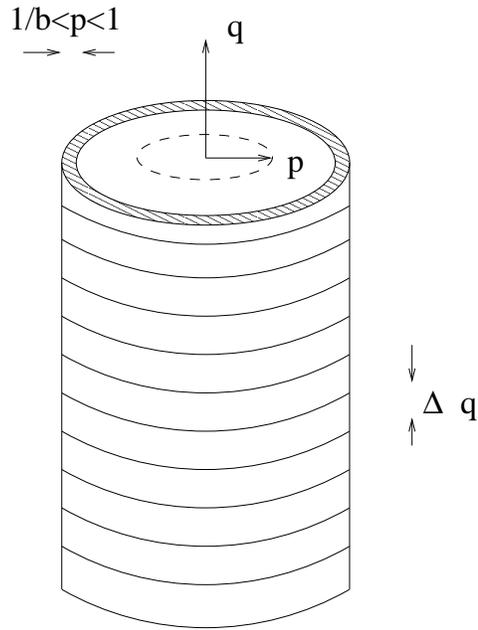}
  \caption{\label{fig1} The modes being integrated out during one 
                        renormalisation step.}
\end{center}
\end{figure}

In order to accomodate all these requirements, we apply a variant of Wilson's 
momentum shell renormalisation method, similar to the one used previously by 
Chakravarty, Halperin, and Nelson for the crossover theory of a quantum 
non-linear sigma model \cite{Chakravarty1}.
(Certainly to one-loop order, the ensuing wave vector shell integrals over 
${\bi p}$, followed by the summations over $q$, are technically considerably 
easier to perform in this approach as compared to the field theory version.)
One renormalisation transformation now consists of two steps: 
Starting from the effective hamiltonian functional ${\cal H}$, we first 
integrate out the short-wavelength order parameter fluctuations in the 
partition function $Z({\cal H})$ in the $d$ directions of the plane, and then
sum over the wave vector components perpendicular to the plane (without any 
cutoff). 
In  wave vector space, the modes to be integrated out are depicted in 
figure~\ref{fig1}.
In the second step, the wave vectors ${\bi k}$, spin variables $S_{\bi{k}}$, 
and system thickness $L$ are rescaled with an as yet arbitrary scale factor 
$b > 1$, and a spin rescaling factor $\zeta(b)$ which eventually has to be 
chosen appropriately in terms of $b$:
\begin{eqnarray}
\label{vt1}
\bi{k}'&=& b \, \bi{k} \ , \\ \label{vt12}
S_{\bi{k}'}'&=&\zeta(b)^{-1} \, S_{\bi{k}} \ , \\ \label{vt3}
L'&=&b^{-1} \, L \ . \label{vt4}
\end{eqnarray}
After this transformation, we arrive at a new free energy functional 
${\cal H}'$ of the same form as before, but with renormalised coefficients 
fulfilling $Z({\cal H}) = Z({\cal H}')$. 
By successively evaluating partial partition functions, and subsequently
applying scale transformations, we approach the interesting infrared behaviour
of our theory, and in effect map the perturbationally inaccessible critical
regime onto a region in parameter space where perturbation theory may be 
applied.
As usual the critical regions themselves, characterised by algebraic 
singularities and scale-invariance, are represented by fixed points of the
renormalisation (semi-)group transformation. 

We apply this scheme in the framework of the one-loop approximation, however
refraining from any $\epsilon$ expansion, for in our problem there is no unique
and common upper critical dimension.
We choose our rescaling factors isotropic in contrast to, e.g., 
Ref.~\cite{Folk}; this induces apparently diverging parameters under the 
renormalisation flow \cite{Aharony1}.
Yet these may be readily absorbed into appropriately defined {\em effective 
couplings}, by means of which we can describe the associated crossover features
\cite{Amit2,Frey1,Frey4,Ried1}.
In contrast to Ref.~\cite{Fisher} we define the spin rescaling factor 
$\zeta(b)$ such that the gradient term of the hamiltonian functional remains
unchanged during the renormalisation procedure. 
Again, this helps us capturing the entire crossover regime, while the 
advantages of the scheme in Ref.~\cite{Fisher}, where the term with the lowest
wave vector power is held fixed, are gathered through the construction and
discussion of suitable effective couplings.

\section{Renormalisation group flow equations}
\label{ren}

In this section, we compute the differential renormalisation group flow
equations for a magnetic layer with the easy axis in the plane, a layer with 
easy axis perpendicular to the plane, and a three-dimensional system. 
Our starting points will be the different hamiltonians (\ref{Stompf}), 
(\ref{Wicht}), and (\ref{3dHamilton}), respectively, for these systems. 
Following through the renormalisation scheme as described in the previous 
section above, with $b^\ell = e^{\delta \ell}$, $\delta \to 0$, for the finite
systems we arrive at the following differential flow equations for the system 
thickness $L$, the coefficient $c$ of the gradient term, and the strength $g$ 
of the dipole-dipole interaction:
\begin{eqnarray}
\label{Lflow}
\frac{\rmd L}{\rmd \ell} &=& - L \ , \\
\label{cflow}
\frac{\rmd c}{\rmd \ell} &=& 0 \ , \\
\label{gflow}
\frac{\rmd g}{\rmd \ell} &=& g \ .
\end{eqnarray}
Before moving on to the flow equations specific to the different systems, we 
discuss these differential equations common to both finite systems. 
The flow equation ({\ref{Lflow}) for the layer thickness $L$ is {\it exact},
and is solved by $L = L_0 \rme^{-\ell}$ with $\ell \to +\infty$ in the infrared
limit. 
Thus, the fixed points for the layer thickness are $L=\infty$ and $L=0$, with 
$L=0$ being infrared-stable. 
Hence the asymptotic critical behaviour is that of the two-dimensional system:
As the correlation length diverges, the system thickness becomes irrelevant in
the renormalisation group sense (as indicated by the negative sign on the 
r.h.s. in equation (\ref{Lflow}).)
The coefficient $c$ of the gradient term in the free energy functional is held
fixed during the renormalisation procedure through the choice of the 
spin rescaling factor $\zeta$, whence equation (\ref{cflow}) follows from 
definition.
The strength of the non-analytic dipole-dipole interaction $g$ cannot be
renormalised perturbationally, and consequently equation (\ref{gflow}) simply 
follows from dimensional analysis and holds to all orders in perturbation
theory.
Its solution reads $g=g_0 \, \rme^{\ell}$, and the fixed points for the dipole 
strength are $g=0$ and $g=\infty$, and the latter is infrared-stable.
As to be expected, the critical behaviour of the system is eventually dominated
by the dipole interaction. 
This is a consequence of the long-range character of the dipolar forces in
contrast to the short-ranged exchange interactions.
For the derivation of the remaining flow equations we have to distinguish 
between the different systems under consideration.

First we focus on the layer with the easy axis in the plane.
The remaining differential flow equations read
\begin{eqnarray}
\frac{\rmd f_1}{\rmd\ell} & = & 0 \ , \\
\frac{\rmd f_2}{\rmd\ell} &=& - 2 \, f_2 \ , \\
\frac{\rmd f_3}{\rmd\ell} &=& - f_3 \ , \\
\frac{\rmd f_4}{\rmd\ell} &=& - 2 \, f_4 \ , \\
\frac{\rmd f_5}{\rmd\ell} &=& - 3 \, f_5 \ ,
\end{eqnarray}
following, to one-loop order, simply from dimensional analysis, and
\begin{eqnarray}
\label{schichtparflow1}
\nonumber\\
\frac{\rmd r}{\rmd\ell} &=& 2 r + \frac{u}{2} \, \frac{1}{(2\pi)^d} \,
\frac{\Lambda^2}{L} \int_0^{2\pi} \! d\theta \ 
\frac{L / 2}{\sqrt{a(\theta)b(\theta)}} \, 
\coth\!\left(\frac{L}{2}\sqrt{\frac{a(\theta)}{b(\theta)}}\right) \ , \\
\label{schichtparflow2} \nonumber \\
\frac{\rmd u}{\rmd \ell} &=& (3-d) \, u - \frac{3}{2} u^2 \, \frac{1}{(2\pi)^d}
\frac{\Lambda^2}{L} \int_0^{2\pi} \! d\theta \  \frac{L/4}{b(\theta)^{1/2} \, 
a(\theta)^{3/2}} \, 
\frac{1}{\sinh(\frac{L}{2}\sqrt{\frac{a(\theta)}{b(\theta)}})} \nonumber \\
&&\qquad \times \left[ \cosh\left(\frac{L}{2}\sqrt{\frac{a(\theta)}{b(\theta)}}
\right) +\frac{L}{2}\sqrt{\frac{a(\theta)}{b(\theta)}} 
\frac{1}{\sinh(\frac{L}{2}\sqrt{\frac{a(\theta)}{b(\theta)}})} \right] \ ,
\end{eqnarray}
to one-loop order.
Here $\theta$ indicates the angle between the wave vector component in the 
plane $\bi{p}$ and the direction of the easy axis $p_l$ 
(see figure~\ref{fig2}).
The $\theta$-dependent functions $a$ and $b$ are defined as follows:
\begin{eqnarray}
a(\theta)&=& r+c\Lambda^2+g(1+f_1) \Lambda \cos^2\theta +gf_3 \Lambda^2
\cos^2\theta +gf_4 \Lambda^3 \cos^2\theta\label{a1} \ , \\
b(\theta)&=&c+gf_2 \Lambda \cos^2\theta+gf_5 \Lambda^2\cos^2\theta\label{b1}\ .
\end{eqnarray}
\begin{figure}[t]
\begin{center}
  \psfrag{q}{$q$}
  \psfrag{p}{${\bi p}$}
  \psfrag{pp}{$p_{l}$}
  \psfrag{theta}{$\theta$}
\includegraphics[width=0.3\textwidth]{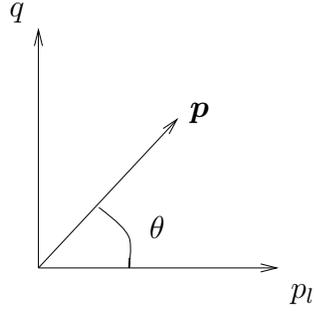}
  \caption{\label{fig2} Definitions of $p_{l}$ and the angle $\theta$. }
\end{center}
\end{figure}
While the coefficient $f_1$ remains invariant under the renormalisation group
flow, the irrelevant parameters $f_2, f_3, f_4$ trivially flow towards their 
stable fixed points $f_2 = f_3 = f_4 = 0$. 
Notice that there exist no fixed points with $f_i=\infty$, because the 
functions $f_i$ were only defined for a finite system thickness. 
The more intricate flow equations for the temperature variable $r$ and the 
non-linear coupling $u$ will be discussed in the following sections.

For a layer with easy axis perpendicular to the plane, on the other hand, we 
find the following flow equations
\begin{eqnarray}
\frac{\rmd \tilde{f}_1}{\rmd\ell} &= & 0 \ , \label{schichtflowsenk4}\\
\frac{\rmd \tilde{f}_2}{\rmd\ell} &=& - 2 \, \tilde{f}_2 \ , 
\label{schichtflowsenk5}\\
\frac{\rmd \tilde{f}_3}{\rmd\ell} &=& - \tilde{f}_3 \ , 
\label{schichtflowsenk7}
\end{eqnarray}
from power counting, and
\begin{eqnarray}
\frac{\rmd r}{\rmd\ell} &=& 2 \, r + \frac{u}{2} \, \frac{K}{L} \sum_{\bi q}
\frac{1}{a+b \, q^2} \ , \label{schichtflowsenk8}\\ 
\frac{\rmd u}{\rmd\ell} &=& (3-d) \, u - \frac{3}{2}u^2 \frac{K}{L} 
\sum_{\bi q} \frac{1}{(a+bq^2)^2} \ . \label{schichtflowsenk9}
\end{eqnarray}
Here the functions $a$ and $b$ are defined as follows
\begin{eqnarray}
a &=& r + (c-g\tilde{f}_3) \, \Lambda^2 - g (1+\tilde{f}_1) \, \Lambda \ , 
\label{adef}\\
b &=& c-g\tilde{f}_2 \, \Lambda-g\tilde{f}_4 \, \Lambda^2 \ ,
\end{eqnarray}
and we have introduced the used the dimension-dependent factor
\begin{equation}
K = \frac{\Lambda^d}{(2 \pi)^{d-1}} \ .
\label{3.76}
\end{equation}
Closer inspection reveals that in the case of an easy axis perpendicular to the
layer and with infinite anisotropy, the propagator is not positive definite. 
Inevitably therefore, the terms in the sums in equations 
(\ref{schichtflowsenk8}) and (\ref{schichtflowsenk9}) will become negative
under the renormalisation group flow.
This indicates an instability at non-zero wave vector, and the ground state of 
the system becomes inhomogeneous, with the characteristic wavelength of the
emerging spatial structures given by that instability wave vector (the typical 
size of the structures appearing in the low-temperature phase is estimated in
appendix B). 
At any rate, it is impossible to analyse the asymptotic critical behaviour 
with the effective hamiltonian (\ref{Wicht}) which was obtained via an 
expansion about a presumed homogeneous ground state. 
Hence, in the remainder of this paper we shall concentrate on the study of the 
critical behaviour of a layer with in-plane easy axis, and on the 
three-dimensional limit.  

For a three-dimensional system, finally we obtain the renormalisation group 
flow equations
\begin{eqnarray}
\frac{\rmd c}{\rmd\ell} &=& 0 \ , \\
\frac{\rmd g}{\rmd\ell} &=& 2 \, g \ , \\
\label{flow3dd1}
\frac{\rmd r}{\rmd\ell} &=& 2 \, r + \frac{u}{4} \, 
\frac{K}{\sqrt{\left(c+\frac{g}{\Lambda^2}\right)\left(r+c \Lambda^2\right)}} 
\ , \\ \label{flow3dd2}
\frac{\rmd u}{\rmd\ell} &=& (4-D) \, u - \frac{3}{8} u^2 \,
\frac{K}{\sqrt{\left(c+\frac{g}{\Lambda^2}\right) \left(r+c\Lambda^2\right)^3}}
\ .
\end{eqnarray}
These equations do not appear as the limit $L \rightarrow \infty$ of the flow
equations of the layer (\ref{schichtparflow1}) and (\ref{schichtparflow2}),
where the effective free energy has already been adjusted to the specific
geometries.
In order to arrive at the correct 3-D expression, one has to start from the 
correct, full hamiltonian (\ref{3dHamilton}), as the long-wavelength limit and
the limit $N \to \infty$ do not commute, as already mentioned in section 
\ref{Propagator}.
Again, $c$ is held fixed by construction, and the renormalisation group flow
for the relevant dipole-dipole interaction has the unstable fixed point $g=0$ 
and the infrared-stable fixed point $g=\infty$. 

Before discussing the different critical regions and the crossover between them
within our scheme, we focus on the obvious limiting cases of the general flow 
equations, and compute the critical exponents within dimensional $\epsilon$ 
expansions about the appropriate upper critical dimensions, respectively.

\section{Limiting cases in $\epsilon$ expansions}
\label{Limit}
\subsection{Three-dimensional Ising limit}

We arrive at the limit of the flow equations at the {\it unstable} fixed point 
$g=0$ and $L=\infty$ by expanding the hyperbolic functions in the differential
flow equations (\ref{schichtparflow1}) and (\ref{schichtparflow2}) for the 
layer in the limit of large argument. 
Assuringly, we can alternatively take the flow equations for the 
three-dimensional system (\ref{flow3dd1}) and (\ref{flow3dd2}) with $g=0$. 
With either approach, we obtain
\begin{eqnarray}
\frac{\rmd r}{\rmd\ell} &=& 2 \, r + \frac{u}{4} \, 
\frac{K}{\sqrt{c \, (r+c\Lambda^2)}} \label{3dirul1} \ , \\
\frac{\rmd u}{\rmd\ell} &=& \epsilon \, u - \frac{3}{8} u^2 \frac{K}{\sqrt{c \,
(r+c\Lambda^2)^3}} \ , \label{3dirul2}
\end{eqnarray}
where $\epsilon=4-D$ gives the distance to the upper critical dimension $D_c=4$
of the system. 
From these equations we get the critical exponents at the non-trivial fixed 
point to first order in $\epsilon$, 
\begin{eqnarray}
r^* &= & -\frac{\epsilon}{3} \, c \, \Lambda^2 + {\cal O}(\epsilon^2) \ , \\
K \, u^* &=& \frac{8}{3} \, \epsilon \, 
\frac{1}{\sqrt{c \, (c \, \Lambda^2)^3}} +{\cal O} (\epsilon^2) \ ,
\end{eqnarray}
as usual via the eigenvalues of the linearised flow equations. 
The results, consistently expanded to first order in $\epsilon = 4 - D$ are
\begin{eqnarray}
\nu &=& \frac{1}{2} \left(1+\frac{\epsilon}{6} + {\cal O}(\epsilon^2) \right) \
, \\ 
\eta &=& 0 + {\cal O}(\epsilon^2) \ , \\
\gamma &=& 1 + \frac{\epsilon}{6} + {\cal O}(\epsilon^2) \ ,
\end{eqnarray}
in accordance with the well-known results for the single-component $\phi^4$
model (see, e.g., Ref.~\cite{Amit1}), with $\epsilon = 1$ in $D = d + 1 = 3$
dimensions.

\subsection{Three-dimensional uniaxial dipolar limit}

Next, we discuss the second limiting case in three dimensions, namely the 
uniaxial dipolar fixed point at $g=\infty$ and $L=\infty$. 
We have to start from equations (\ref{flow3dd1}) and (\ref{flow3dd2}). 
Upon introducing the effective expansion parameter \cite{Frey1}
\begin{eqnarray}
v_{\rm eff} =  u / \sqrt{g} \ , 
\end{eqnarray}
we arrive at the flow equations
\begin{eqnarray}
\frac{\rmd r}{\rmd\ell} &=& 2 \, r + \frac{v_{\rm eff}}{4} \, 
\frac{K \, \Lambda}{\sqrt{r + c \, \Lambda^2}} \ , \\
\frac{\rmd v_{\rm eff}}{\rmd\ell} &=& \epsilon \, v_{\rm eff} - \frac{3}{8} \, 
v_{eff}^2 \, \frac{K \, \Lambda}{(r+c \Lambda^2)^{3/2}} \ ,
\end{eqnarray}
where now $\epsilon=3-D$ was used for the deviation from the upper critical
dimension $D_c = 3$. 
In $D = d + 1 = 3$ dimensions, we thus expect the mean-field critical exponents
with logarithmic corrections. 
This is in agreement with previous work addressing the critical behaviour of a
three-dimensional uniaxial dipolar magnet \cite{Larkin,Frey1,Brezin}.

\subsection{Two-dimensional Ising limit}

In the limit of absent dipole-dipole interaction $g=0$ and zero thickness 
$L=0$ we have to expand the hyperbolic functions in the flow equations 
(\ref{schichtparflow1}) and (\ref{schichtparflow2}) for vanishing arguments. 
With the effective coupling
\begin{equation}
v_{\rm eff} = u / L \ ,
\end{equation}
this procedure gives the flow equations
\begin{eqnarray}
\frac{\rmd r}{\rmd\ell} &=& 2 \, r + \frac{v_{\rm eff}}{2} \, 
\frac{K}{(r + c \, \Lambda^2)} \ , \\
\frac{\rmd v_{\rm eff}}{\rmd\ell} &=& \epsilon \, v_{\rm eff} - \frac{3}{2} \,
v_{eff}^2 \, \frac{K}{(r + c \, \Lambda^2)^2} \ ;
\end{eqnarray}
here, however, $\epsilon = 4 - d$ $(=2$ in $D = d + 1 = 3$ dimensions).
Linearising in the vicinity of the non-trivial fixed point 
\begin{eqnarray}
r^* &=& - \frac{\epsilon}{6} \, c \, \Lambda^2 + {\cal O}(\epsilon^2) \ , \\
K \, v_{\rm eff}^* &=& \frac{2}{3} \, \epsilon \, c^2 \, \Lambda^4 + 
{\cal O}(\epsilon^2) \ ,
\end{eqnarray}
we find the critical exponents to order $\epsilon=4-d(=2$ in two dimensions):
\begin{eqnarray}
\nu &=& \frac{1}{2} \left( 1 + \frac{\epsilon}{6} + {\cal O}(\epsilon^2)
\right) \ ,\\ 
\eta &=& 0 + {\cal O}(\epsilon^2) \, \\
\gamma &=& 1 + \frac{\epsilon}{6} + {\cal O}(\epsilon^2) \ .
\end{eqnarray}
Thus, this low-dimensional limit is also correctly included in the general flow
equations. 
Note that in terms of the expansion parameter $\epsilon$, these exponents of 
course coincide with the three-dimensional Ising case above.
Remember, however, that there $\epsilon = 4 - D = 1$ and here 
$\epsilon = 4 - d =2$ in the relevant physical situation $D = d + 1 = 3$.

\subsection{Two-dimensional uniaxial dipolar limit, in-plane easy axis}

This limit is especially interesting for two reasons: First, it represents the
asymptotic critical behaviour of the system under consideration; and second it 
is intriguing on its own right because we are aware of only one other reference
which addresses the critical behaviour of a two-dimensional uniaxial dipolar 
system with the easy axis in the plane \cite{DeBell}. 
Yet, in Ref.~\cite{DeBell} the critical exponents are not presented explicitly.
Using the same approximation for the hyperbolic functions as before and 
introducing the effective expansion parameter 
\begin{eqnarray}
v_{\rm eff}=\frac{u}{L \, \sqrt{g}} \ ,
\end{eqnarray}
we obtain the asymptotic flow equations
\begin{eqnarray}
\frac{\rmd r}{\rmd\ell} &= & 2 \, r + \frac{v_{\rm eff}}{2^d \pi^{d-1}} \, 
\frac{\Lambda^2}{\sqrt{(r + c \, \Lambda^2) \, (1+f_1) \, \Lambda}} 
\label{4.45} \ , \\
\frac{\rmd v_{eff}}{\rmd\ell} &=& \left( \frac{7}{2}-d \right) \, v_{eff} -
\frac{3}{2} \, \frac{v_{\rm eff}^2}{2^d \pi^{d-1}} \, 
\frac{\Lambda^2}{\sqrt{(r + c \, \Lambda^2)^3 \, (1+f_1) \, \Lambda}} \ .
\label{4.46}
\end{eqnarray}
Consequently, we identify the upper critical dimension here as $d_c = 7/2$.
The non-trivial fixed point to order $\epsilon = \frac{7}{2}-d$ is 
\begin{eqnarray}
r^* &=& - \frac{\epsilon}{3} c \, \Lambda^2 +{\cal O}(\epsilon^2)\ , \\ 
v_{\rm eff}^* &=& \frac{2}{3} \, \epsilon \frac{2^d \pi^{d-1}}{\Lambda^2} 
\left[(1+f_1) \Lambda \right]^{1/2} (c \, \Lambda^2)^{3/2} +{\cal
O}(\epsilon^2) \ .
\end{eqnarray}
Linearisation near this fixed point leads to the following asymptotic critical 
exponents to first order in $\epsilon=\frac{7}{2}-d$
\begin{eqnarray}
\nu &=& \frac{1}{2} \left( 1 + \frac{\epsilon}{6} + {\cal O}(\epsilon^2)
\right) \ ,\\ 
\eta &=& 0 + {\cal O}(\epsilon^2)\ , \\
\gamma &=& 1 + \frac{\epsilon}{6} + {\cal O}(\epsilon^2) \ .
\end{eqnarray}
Once again, these exponents look superficially identical to those of the 
short-range Ising model; yet the critical dimension and the value of $\epsilon$
is different: in $d = 2$, $\epsilon = 1/2$.
In summary, all the asymptotic limits are correctly included in the general 
flow equations.

\section{Crossover analysis}
\label{crossover}

In the previous section, we have demonstrated the consistency of our general
approach with the $\epsilon$ expansions near the respective critical dimensions
of the limiting universality classes.
We now proceed to the application of our scheme to the description of the 
actual crossovers between the different critical regions.

\subsection{General procedure}

At first we want to describe the general procedure to be used. 
We use the calculated renormalisation group flow equations to investigate the
scale dependence of the effective interaction parameters.
Specifically, we map the critical region onto a region in parameter space that 
is perturbationally accessible.
In the infrared region, we can, e.g., evaluate the static susceptibility 
perturbationally, and thus obtain the static susceptibility $\chi$ in the
critical region via the solution of the renormalisation group equation that
takes the form (the Fisher exponent $\eta$ vanishes in the one-loop 
approximation)
\begin{eqnarray}
\chi(r_0,g_0,u_0,L_0,\bi{k}) = \rme^{2 \ell} \
\chi\Bigl(r(\ell),g(\ell),u(\ell),L(\ell),\bi{k}(\ell)\Bigr) \ .
\end{eqnarray}
Therefrom we can infer the {\it effective} exponent 
\begin{eqnarray}
\gamma_{\rm eff}(\tau) = \frac{\rmd \ln \chi^{-1}}{\rmd\ln \tau}
\end{eqnarray}
as function of reduced temperature $\tau = \frac{T-T_c}{T_c}$.
This method was used first in \cite{Rudnick1} in the framework of an $\epsilon$
expansion; but here we make no use of any dimensional expansion, but perform 
all all our calculations numerically at a {\it fixed} dimension. 
The one-loop recursion relations are then solved without recourse to any 
additional approximation. 

It is vital here to introduce an appropriate effective expansion parameter that
yields the correct 3-D and 2-d limits as given in the previous section. 
This effective coupling has to have a finite limit at the asymptotic fixed 
point below the upper critical dimension of the system.
As suggested by the flow equations in section \ref{ren}, we use the effective 
couplings 
\begin{eqnarray}
v_{\rm eff} = u \int_0^{2\pi} \! d\theta \  \frac{1}{\sqrt{a(\theta)
b(\theta)}} \, 
\coth\!\left(\frac{L}{2} \sqrt{\frac{a(\theta)}{b(\theta)}}\right)
\end{eqnarray} 
for the layer with in-plane easy axis, and
\begin{eqnarray}
v_{\rm eff} = \frac{2\pi}{\sqrt{r+c\Lambda^2}} \frac{u}{\sqrt{g}}
\end{eqnarray}
for the three-dimensional case.
Then within our methodical framework, the critical regions are again 
represented by fixed points of the renormalisation group transformation, and we
may obtain the critical exponents from the eigenvalues of the linearised 
transformation in the vicinity of the fixed points (yet without applying any 
further approximation like dimensional expansions).

\subsection{Limiting cases at fixed dimension}

The appearance of renormalisation group fixed points to first order in an
expansion does not necessarily guarantee the existence of fixed points if we 
refrain from an $\epsilon$ expansion.
At first, therefore, we have to re-calculate the non-trivial fixed points
and the associated critical exponents at fixed dimensionality $D = d + 1$. 
The actual fixed points thus obtained are listed in table \ref{fptable}.
\begin{table}
\caption{\label{fptable} Renormalisation group fixed points at fixed dimension
        to one-loop order.}
\begin{tabular}{@{}lll}
\br
Fixed point& $r^*$ & $v_{\rm eff}^*$ \\ 
\mr
3-D Ising & $-\frac{2}{3}(4-D)c\Lambda^{d}\left[2+\frac{2}{3}(4-D)\right]^{-1}$
& $\frac{16 \pi}{3 K_d}(4-D) \sqrt{(r^*+c\Lambda^2)}$\\ 
2-d Ising& $-(4-d) c \Lambda^2/(4-d+6)$&
$\frac{16 \pi}{K_d} (4-d) c \Lambda^2/ (4-d+6)$ \\ 
3-D uniaxial dipolar & $0$& $0$ \\
2-d uniaxial dipolar & $-\left( \frac{7}{2}-d \right) c \Lambda^2 / \left(
\frac{7}{2}-d+3 \right)$&
$\frac{16 \pi}{K_d} \left( \frac{7}{2}-d \right) c \Lambda^2 / \left(
\frac{7}{2}-d+3 \right)$\\
\br
\end{tabular}
\end{table}
The eigenvalues of the linearised flow equations can be easily calculated 
analytically, and are not presented explicitly here. 
In table~\ref{Tab5.1} we give the numerical values of the fixed points as well 
as the critical exponents $\nu$ and $\gamma$ for the system with two infinite 
dimensions ($D = d + 1 = 3$).
\begin{table}
\caption{\label{Tab5.1} Numerically determined fixed point values and critical
exponents.}
\begin{indented}
\item
\begin{tabular}{@{}llllll}
\br
Fixed point& $r^*$ & $v_{\rm eff}^*$ & $L^*$ & $\gamma$ &$\nu$ \\
\mr
3-D Ising & $-0.25$&
$78.95$ &$\infty$&$1$&$1/2$\\ 
2-d Ising& $-0.25$&
$78.95$ &$0$&$1$&$1/2$\\
3-D uniaxial dipolar& $0$&
$0$ &$\infty $&$1$&$1/2$\\
2-d uniaxial dipolar & $-0.33$&
$105.27$ &$0$&$0.94$&$0.47$\\
\br
\end{tabular}
\end{indented}
\end{table}
As is readily seen, the numerical results for the critical exponents cannot be
judged as good approximations to their realistic values. 
The reason is very likely the inadequacy of the one-loop approximation. 
An unfortunate artefact of the system under consideration is that the actual
fixed point values as well as the critical exponents for the 3-D Ising and the 
2-d Ising limit turn out identical to this order. 
This happens to be the case {\it only} in this dimension $D=3$, as can be seen 
from figure~\ref{nuIsingfig}, where we plot the numerical values of the 
critical exponent $\nu$ at both the higher and lower dimensional fixed points,
respectively, vs the dimension. 
This artificial coincidence only appears at the dimensions $d=\frac{5}{2}$ and 
$d=2$, the physical dimension in which we are interested. 
Nevertheless we shall argue for the usefulness of our scheme in describing 
different crossover features.
\begin{figure}
\begin{center}
  \includegraphics[width=0.5\textwidth]{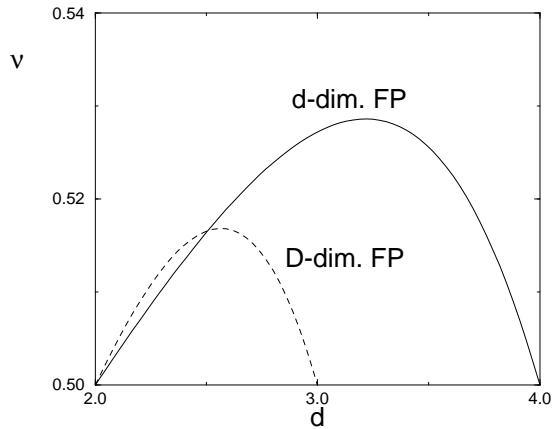}
  \caption{\label{nuIsingfig} Ising fixed point values of the critical exponent
         $\nu$ as function of dimension.}
\end{center}
\end{figure}

\subsection{Dimensional crossover}

The first crossover that we want to analyse within our scheme is the 
dimensional crossover for an Ising layer (dipolar strength $g = 0$).
In order to circumvent the unfortunate exponent coincidence mentioned above, we
evaluate the crossover features between 3.2 and 2.2 dimensions.
In this situation, we are not only able to calculate the flow equations
describing  
the crossover, but also the effective critical exponent $\gamma_{\rm eff}$ as a
function of the reduced temperature $\tau$.
The corresponding renormalisation group fixed point values are listed in 
table~\ref{tab5.2}.
\begin{table}
\caption{\label{tab5.2} Numerical fixed point values for an Ising layer with
        $d=2.2$.} 
\begin{indented}
\item 
\begin{tabular}{@{}lllll}
\br
fixed point& $r^*$ & $v_{\rm eff}^*$ & $L^*$ &$\nu$ \\ 
\mr 
3.2-D Ising & $-0.2105$&
$96.0267$ &$\infty$&$0.5090$\\
2.2-d Ising& $-0.2308$&
$105.2601$ &$0$&$0.5070$\\
\br
\end{tabular}
\end{indented}
\end{table}
The effective exponent $\gamma_{\rm eff}$ describing the crossover is 
calculated numerically for different layer thicknesses $L$, and the
results are plotted in figure~\ref{gamIsingfig}. 
\begin{figure}
\begin{center}
\includegraphics[width=0.8\textwidth]{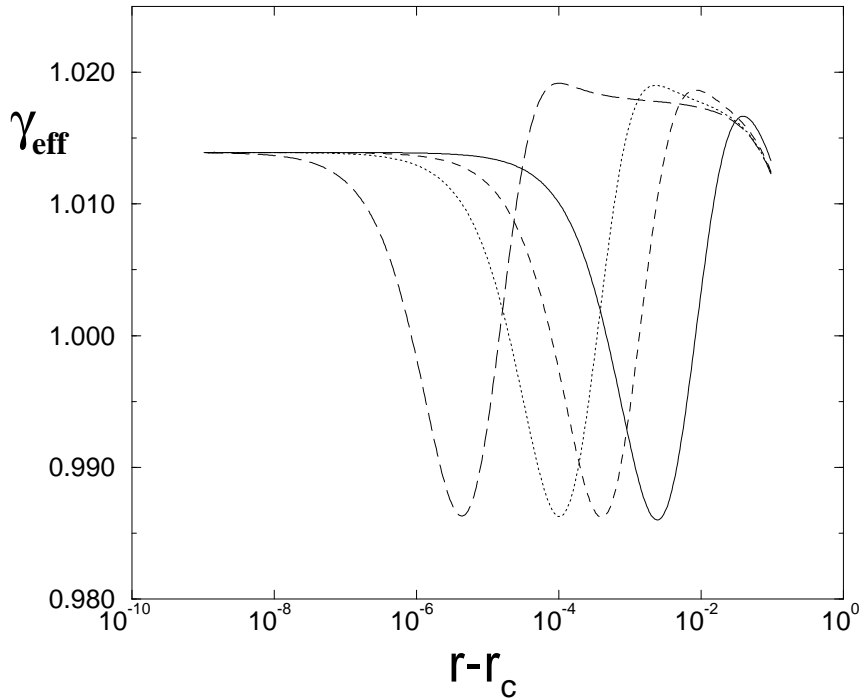}
  \caption{\label{gamIsingfig} Layer with vanishing dipole-dipole interaction:
  Effective critical exponent $\gamma_{\rm eff}$ vs reduced temperature 
  $\tau \sim r - r_c$ for systems of different thicknesses: $N=20$
  (full line), $N=50$ (dashed), $N=100$ (dotted), $N=500$ (long-dashed).}
\end{center}
\end{figure}
It can be seen that at a temperature sufficiently far from the transition
temperature, the susceptibility exponent takes on the value of the 
higher-dimensional fixed point, then with decreasing temperature goes through a
minimum which is obviously independent of the layer thickness, and finally 
reaches the asymptotic value of the lower-dimensional fixed point. 
The thicker the layer, the system remains the longer in the higher-dimensional 
critical region as the temperature is lowered, and reaches the later the 
asymptotic critical region.
This is fully to be expected based on straightforward considerations.

\subsection{Crossover exchange-dominated / dipole-dominated}

Next we apply our scheme to the fully three-dimensional system, where the
crossover occurs between the region with dominating exchange interactions and
the asymptotic region with dominating dipole-dipole interaction. 
Because the numerical values of the critical exponents happen to be identical 
at both these fixed points within our scheme at fixed dimension $D=3$, see
table~\ref{Tab5.1}, the crossover cannot be noticed at all in the temperature 
dependence of the effective critical exponent $\gamma_{\rm eff}$. 
However, we can discuss the crossover by means of the flow diagrams in the 
$r-v_{\rm eff}$ plane. 
In figure~\ref{rv3dfig1}, the flow diagram is depicted for different initial 
temperatures (which cannot be distinguished in this graph, however). 
The arrow indicates the starting point of the trajectories. 
These first approach the unstable Ising fixed point (filled circle on the 
right), before finally approaching the asymptotically stable fixed point that 
represents the uniaxial dipolar regime (filled circle on the left). 
Eventually the trajectories diverge in the temperature variable $r$ due to the 
numerically inevitably finite distance from the critical point (even small
inaccuracies in determining the true critical temperature will eventually blow
up under renormalisation).
\begin{figure}
\begin{center}
\includegraphics[width=0.8\textwidth]{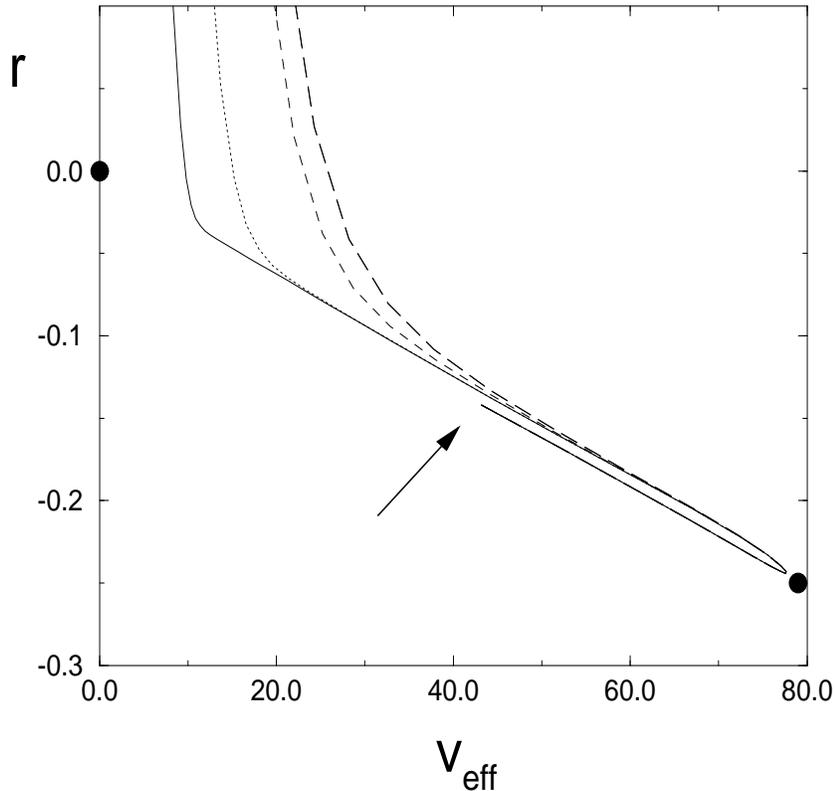}
  \caption{\label{rv3dfig1} Infinite three-dimensional dipolar system: Flow of
  the effective coupling $v_{\rm eff}$ and temperature variable  $r$ for 
  different initial temperature values, with $g_0=10^{-5}$ and $u_0=5$: 
  $r\approx r_c$ (full line), $r-r_c= 6.42 \cdot 10^{-11}$ (dotted),
  $r-r_c = 6.66 \cdot 10^{-9} $ (dashed), $r-r_c=1.67 \cdot 10^{-8}$ 
 (long-dashed). 
}
\end{center}
\end{figure}
In figure~\ref{rv3d2fig} we depict the same scenario for a stronger 
dipole-dipole interaction. 
In this situation, the unstable fixed point is not approached as closely as 
before, and the stable fixed point is approached more closely before the
trajectories eventually diverge to $r \to \infty$ due to the finite distance 
from the critical temperature.
\begin{figure}
\begin{center}
\includegraphics[width=0.8\textwidth]{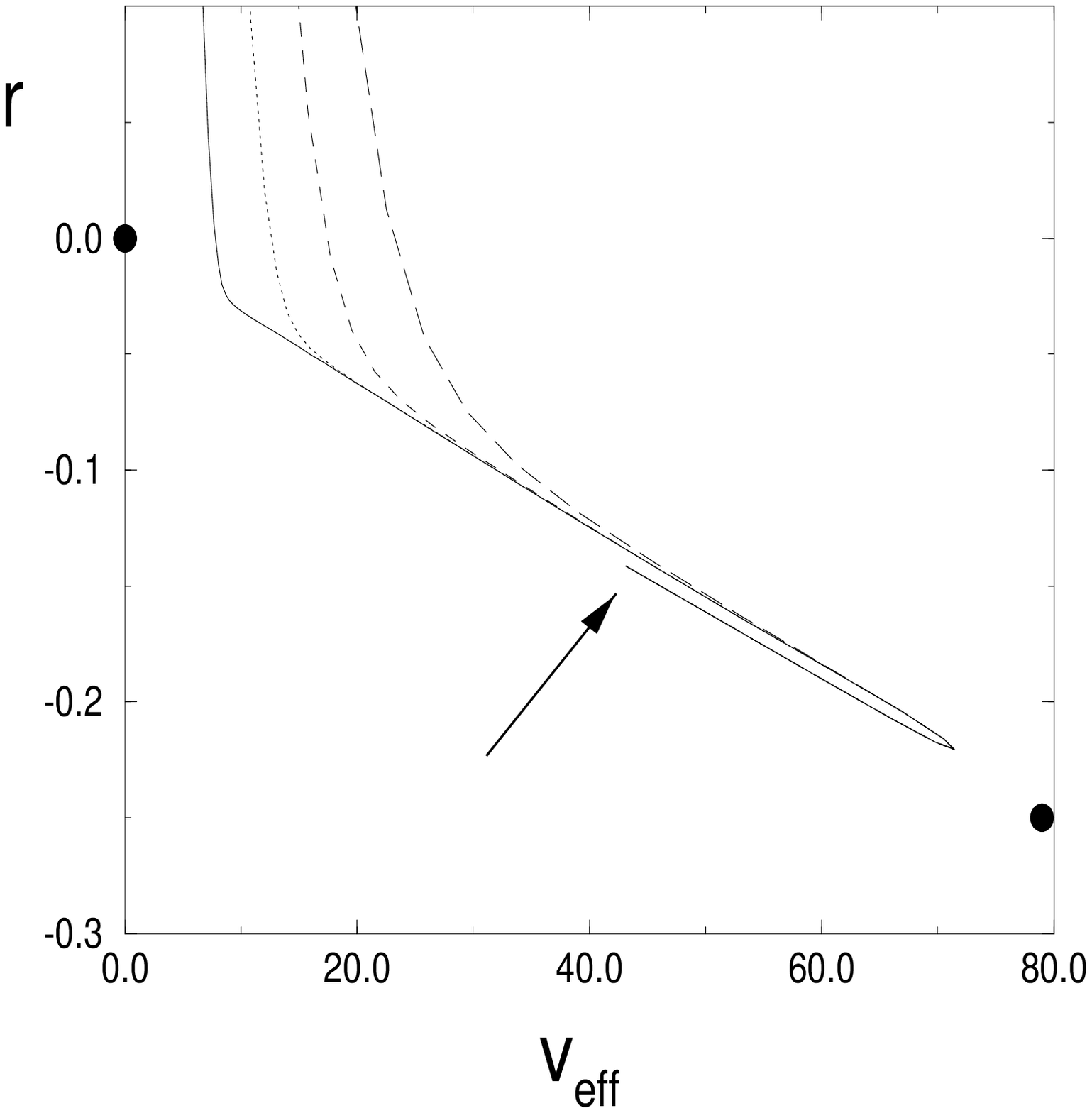}
  \caption{\label{rv3d2fig} Infinite three-dimensional dipolar system: Flow of
  the effective coupling  $v_{\rm eff}$ and temperature variable $r$ for 
  different initial temperature values, with $g_0=10^{-3}$ and $u_0=5$:
  $r\approx r_c$ (full line), $r-r_c=4.44 \cdot 10^{-10}$ (dotted),
  $r-r_c=4.04 \cdot 10^{-8}$ (dashed), $r-r_c=7.40 \cdot 10^{-7}$ 
  (long-dashed).
  }
\end{center}
\end{figure}

\subsection{Crossover features of the dipolar layer}

At last, we address the description of the crossover scenario for a dipolar 
magnetic layer with in-plane easy axis. 
As the starting point is the hamiltonian (\ref{Stompf}), which is only defined 
for finite system thickness, we cannot reach the fixed point corresponding to 
the three-dimensional uniaxial dipolar region. 
The crossover will thus occur between the critical regions of the 
three- and two-dimensional Ising systems, and the two-dimensional uniaxial 
dipolar regime, which represents the asymptotic universality class.
As the numerical fixed point values of the effective coupling, the temperature
variable, and the effective critical exponent of the static susceptibility
$\gamma_{\rm eff}$ of the two- and the three-dimensional Ising fixed point are 
identical, we illustrate the flow trajectories in three-dimensional parameter
space, consisting of $r$, $v_{\rm eff}$, and the thickness-dependent variable 
$\frac{N}{N+1}$ as third coordinate. 
In terms of this quantity, the three- and two-dimensional Ising fixed points
assume the value 1 and 0, respectively, and thus become distinguishable. 
In figures~\ref{rvD3}, \ref{rvD2}, and \ref{rvD5} we show the trajectories with
initial values in the vicinity of the critical temperature for increasing 
values of the dipole strength. 
At first the trajectories approach the three-dimensional Ising fixed point, 
then the two-dimensional Ising fixed point, and finally the asymptotic 
two-dimensional uniaxial dipolar fixed point. 
Eventually even the tiny numerical deviation from the critical temperature 
gives rise to a divergence in $r$ again. 
For stronger dipole interaction, the Ising fixed points are not approached as 
closely. 
\begin{figure}
\begin{center}
   \psfrag{n}{$\frac{N}{N+1}$}
   \psfrag{v}{$v_{eff}$}
   \psfrag{r}{$r$}
   \psfrag{3}{3dI}
   \psfrag{2}{2dI}
   \psfrag{2d}{2dud}
   \includegraphics[width=0.6\textwidth]{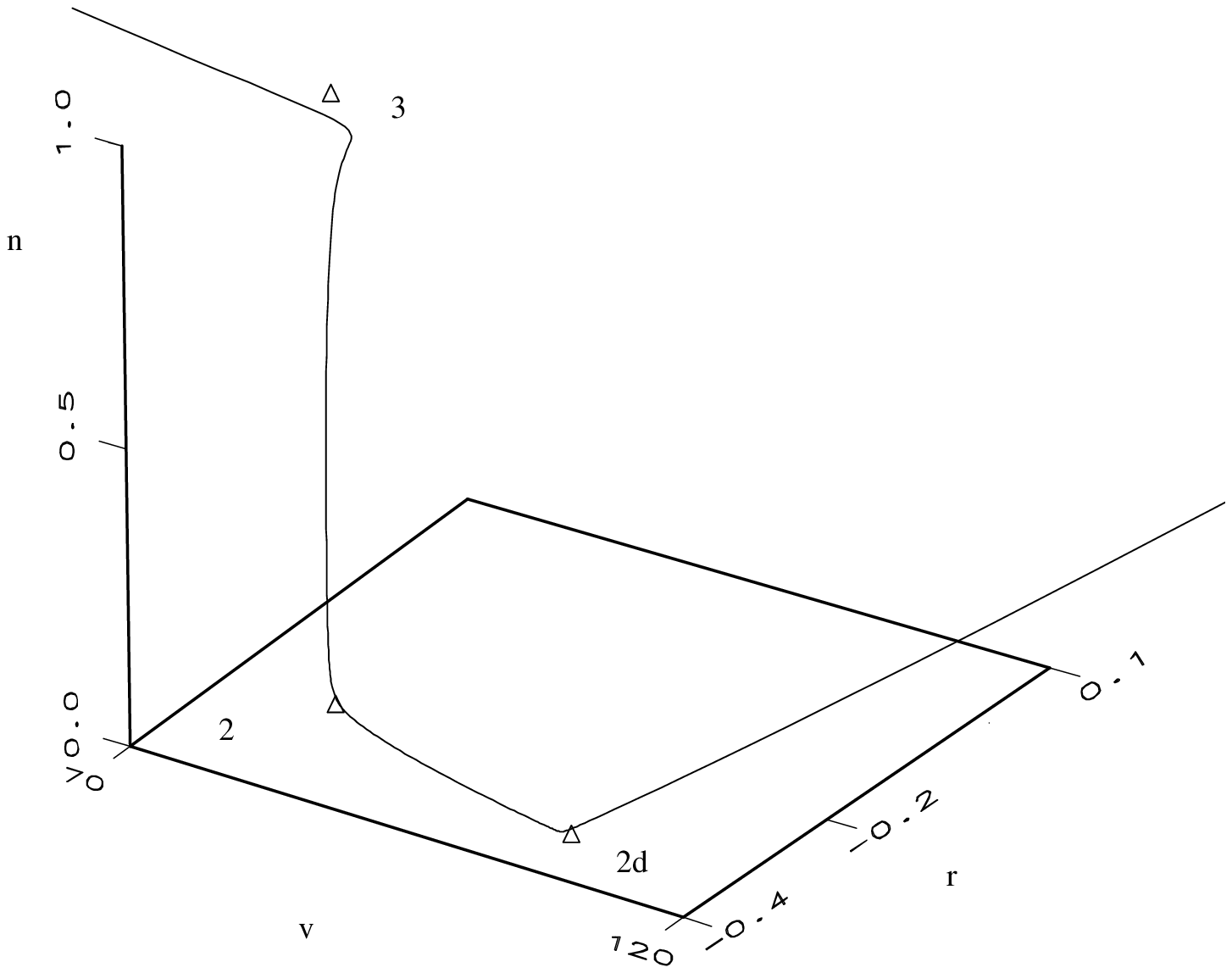}
  \caption{\label{rvD3} Dipolar layer with easy in-plane axis: Flow diagram for
   $g_0=10^{-7}, u_0=10, r_0 \sim r_c,  N_0=100$.} 
\end{center}
\end{figure}
\begin{figure}
\begin{center}
   \psfrag{n}{$\frac{N}{N+1}$}
   \psfrag{v}{$v_{eff}$}
   \psfrag{r}{$r$}
   \psfrag{3}{3dI}
   \psfrag{2}{2dI}
   \psfrag{2d}{2dud}
   \includegraphics[width=0.6\textwidth]{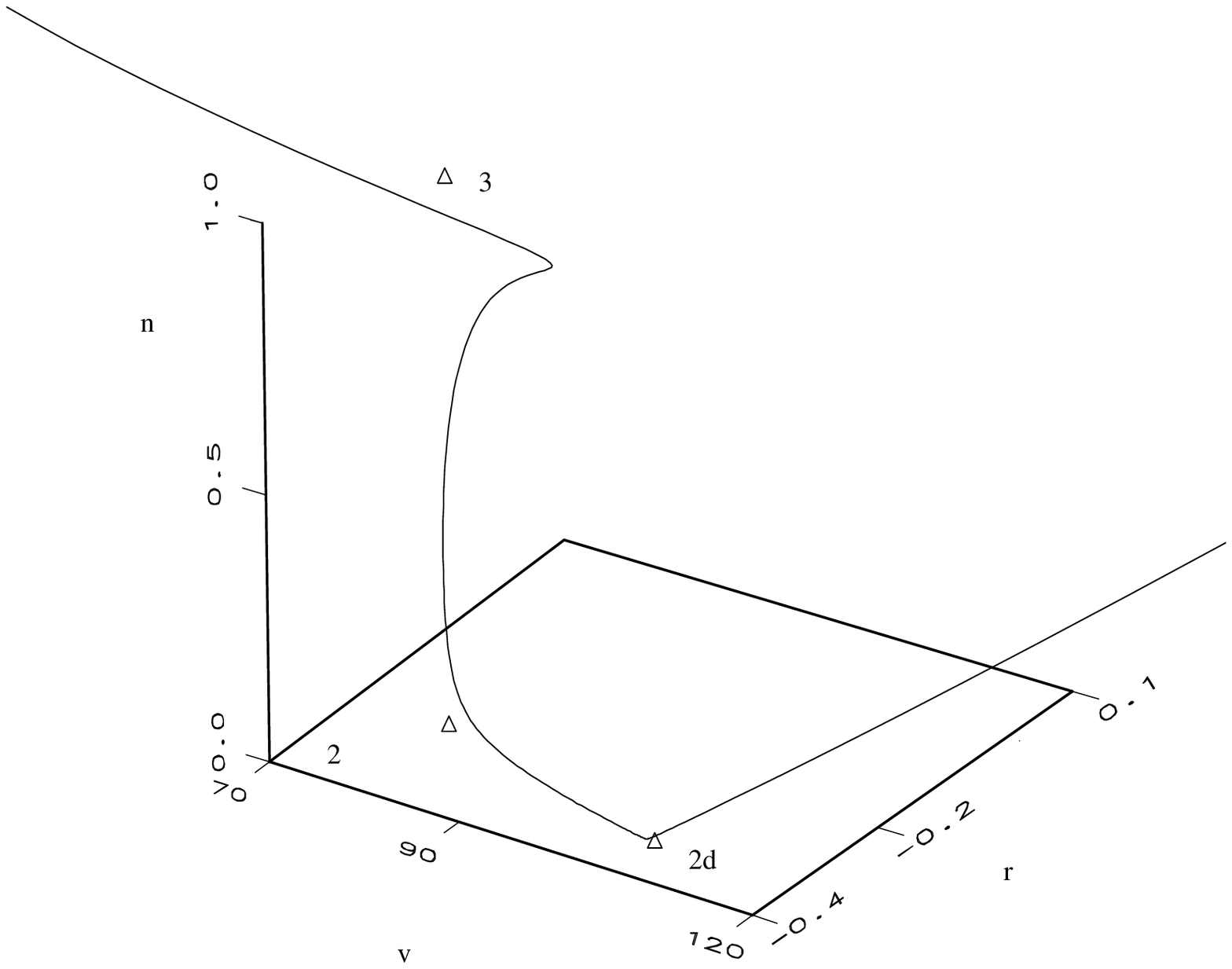}
  \caption{\label{rvD2} Dipolar layer with easy in-plane axis: Flow diagram for
  $g_0=10^{-6}, u_0=10, r_0 \sim r_c, N_0=100$.}
\end{center}
\end{figure}
\begin{figure}
\begin{center}
   \psfrag{n}{$\frac{N}{N+1}$}
   \psfrag{v}{$v_{eff}$}
   \psfrag{r}{$r$}
   \psfrag{3}{3dI}
   \psfrag{2}{2dI}
   \psfrag{2d}{2dud}
   \includegraphics[width=0.6\textwidth]{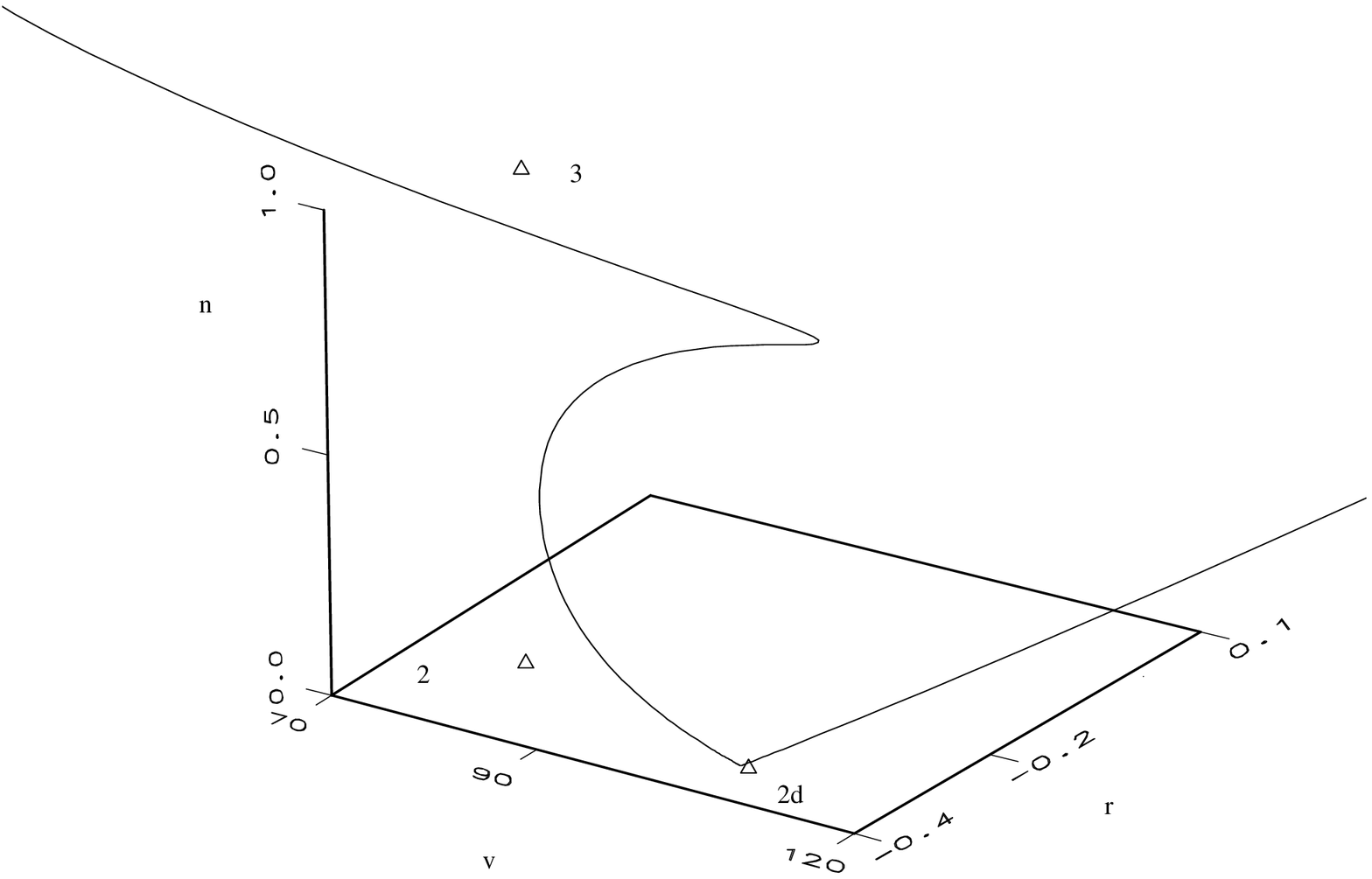}
  \caption{\label{rvD5} Dipolar layer with easy in-plane axis: Flow diagram for
  $g_0=10^{-5}, u_0=10, r_0 \sim r_c,  N_0=100$.}
\end{center}
\end{figure}
In figure~\ref{tempabh} we furthermore depict the temperature dependence of the
trajectories: 
The higher the temperature the sooner the trajectories start to diverge.  
\begin{figure}
\begin{center}
   \psfrag{N}{$\frac{N}{N+1}$}
   \psfrag{v}{$v_{eff}$}
   \psfrag{r}{$r$}
   \psfrag{3}{3dI}
   \psfrag{2}{2dI}
   \psfrag{2d}{2dud}
\includegraphics[width=0.7\textwidth]{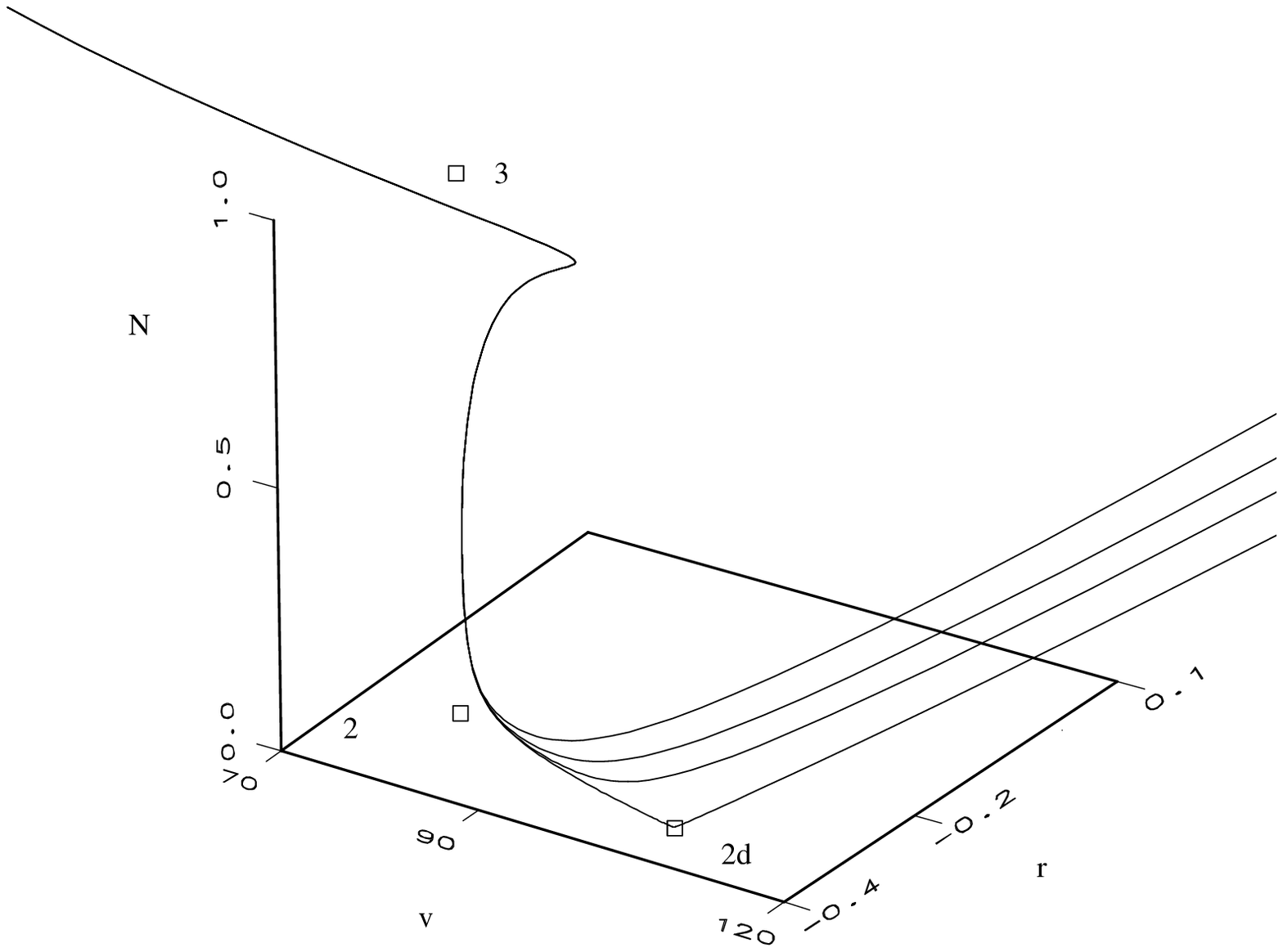}
  \caption{\label{tempabh} Dipolar layer with easy in-plane axis: Flow diagram
  for $g_0=10^{-6}, u_0=10, r_0-r_c=8.33\cdot 10^{-10}, 2.33\cdot 10^{-10}, 
  0.53 \cdot 10^{-10}, r_0 \sim r_c,(\mbox{from left to right}), N_0=100$.}
\end{center} 
\end{figure}

\section{Summary}
\label{summary}

In this paper we have analysed crossover scenarios for a uniaxial dipolar
magnetic layer. 
We have derived the appropriate effective free energy functional in the 
long-wavelength limit for a finite system with thickness $L$ in one spatial 
direction; the resulting dipolar hamiltonian turned out to be considerably 
different from the one for the infinite three-dimensional system. 
In the long-wavelength limit, which does not commute with the limit 
$L \to \infty$, the propagator of the three-dimensional system cannot be 
recovered from the propagator of the layer in the limit of infinite layer 
thickness.

We have presented a unified renormalisation scheme, within which the critical 
behaviour of the layer with finite thickness can nevertheless be treated as 
well as that of the infinite system. 
This renormalisation procedure is required to be capable of capturing the 
crossovers between critical regions with different system dimensions, different
analytical structures of the propagator due to the dipole-dipole interaction,
and consequently different upper critical dimensions. 
Hence we had to refrain from any dimensional $\epsilon$ expansion near any of
the fixed points, and have derived the renormalisation group flow equations in
the one-loop approximation at fixed dimension. 
Furthermore, we have introduced a common effective expansion parameter, as 
suggested by the flow equations, which leads to finite non-trival fixed points 
charactarising the different critical regions. 
As a test of the consistency of our approach, the critical exponents in the 
limiting cases of the 3-d Ising, 2-d Ising, 3-d uniaxial dipolar, and 2-d 
uniaxial (in-plane easy axis) dipolar universality classes were calculated 
within $\epsilon$ expansions with respect to the appropriate upper critical 
dimensions, respectively, and reproduced correctly known results from the 
literature. 
The critical exponents for the asymptotic limit of a two-dimensional uniaxial
dipolar system with in-plane easy axis were, to our knowledge, computed 
explicitly for the first time by means of an expansion near the upper critical
dimension $d_c = 7/2$.

Performing the calculations without the $\epsilon$ expansion, however, reveals 
an unfortunate insufficiency of the one-loop approximation: 
The numerical values of the temperature variable, the effective coupling, and 
the critical exponents at the three-dimensional and the two-dimensional Ising 
fixed points happen to be identical in $D = 3$ dimensions.
In order to test our scheme, we have first evaluated the dimensional crossover 
of the effective critical exponent $\gamma_{\rm eff}$ of an Ising system from 
$D = 3.2$ to $D = 2.2$ dimensions. 
Second, the crossover features of the three-dimensional uniaxial dipolar system
as well as the uniaxial dipolar layer were analysed and discussed in form of 
renormalisation group flow diagrams.
Thus, we hope to have demonstrated the feasibility of the description of such
very complex crossover features within the framework of our renormalisation
group approach, which should be further applicable to a diverse range of 
interesting physical systems.

\ack
We acknowledge support from the German Education and Research Department (BMBF)
under contract no. 03-SC5-TUM 0, and from the Deutsche Forschungsgemeinschaft
(DFG) ,
contract nos. Schw 348/12-1 and Ta 177/2-1,2.

\appendix

\section{Critical temperature}

A number of the above calculations depended on the knowledge of the phase
transition temperature which, due to thermal fluctuations, is shifted downwards
as compared to the mean-field transition temperature. 
The correct location of the critical point (to one-loop order) was determined 
numerically through identifying that starting point of the temperature variable
$r$, which separates those renormalisation group flow trajectories that 
eventually flow to $r=+\infty$ from the ones that eventually flow to 
$r=-\infty$.

Figures~\ref{rc1} and \ref{rc} show the dependence of the numerically obtained
phase transition temperature $r_c$ upon the strength of the dipole-dipole 
interaction $g$ and the layer thickness $N$. 
As one would expect physically, $r_c$ grows both with increasing dipole 
strength and increasing layer thickness.
\begin{figure}
\begin{center}
  \includegraphics[width=0.7\textwidth]{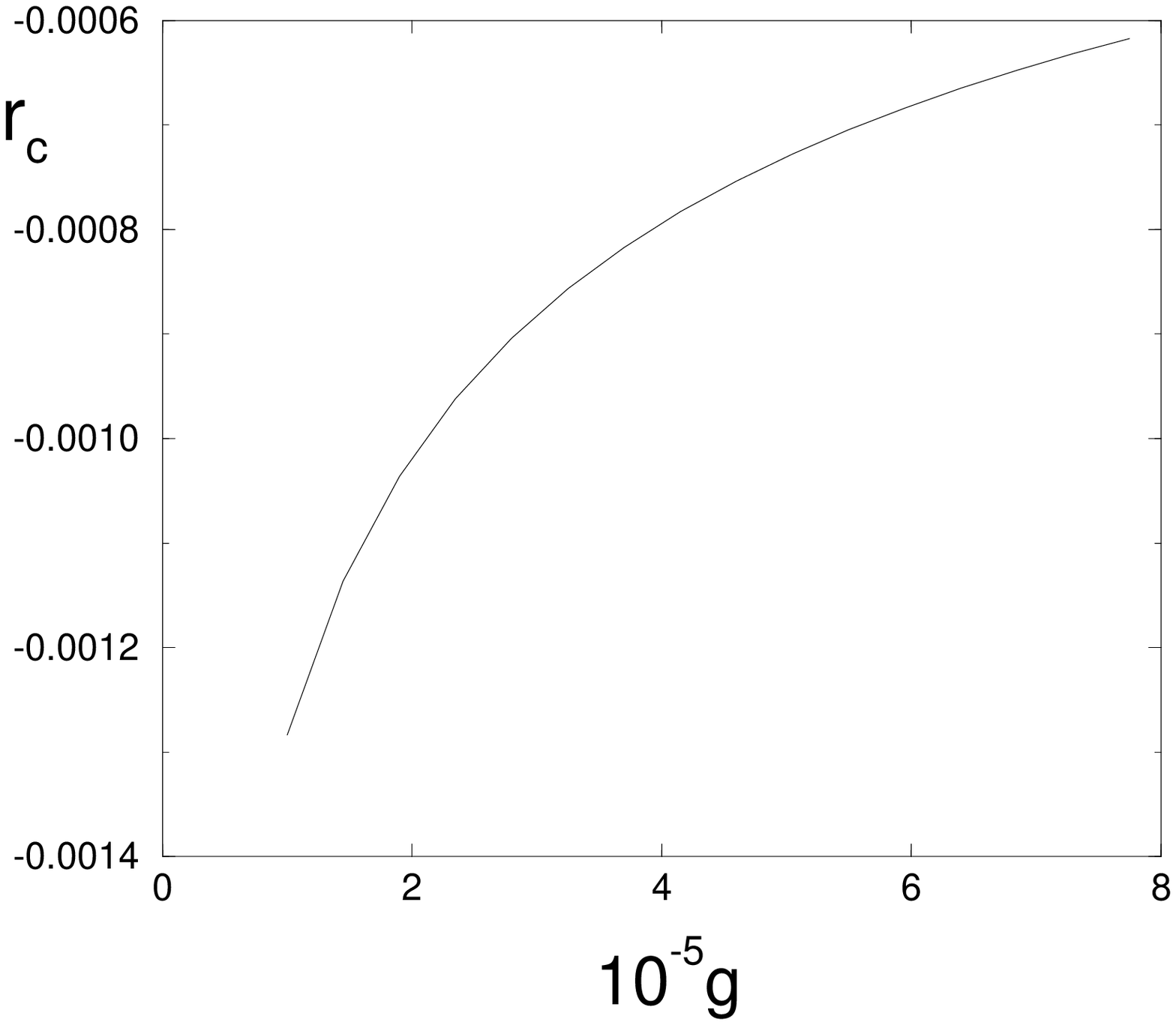}
  \caption{\label{rc1} Phase transition temperature $r_c(g)$ vs dipole strength
  ($N_0=100, u_0=0.1$).}
\end{center}
\end{figure}
\begin{figure}
  \begin{center}
  \includegraphics[width=0.7\textwidth]{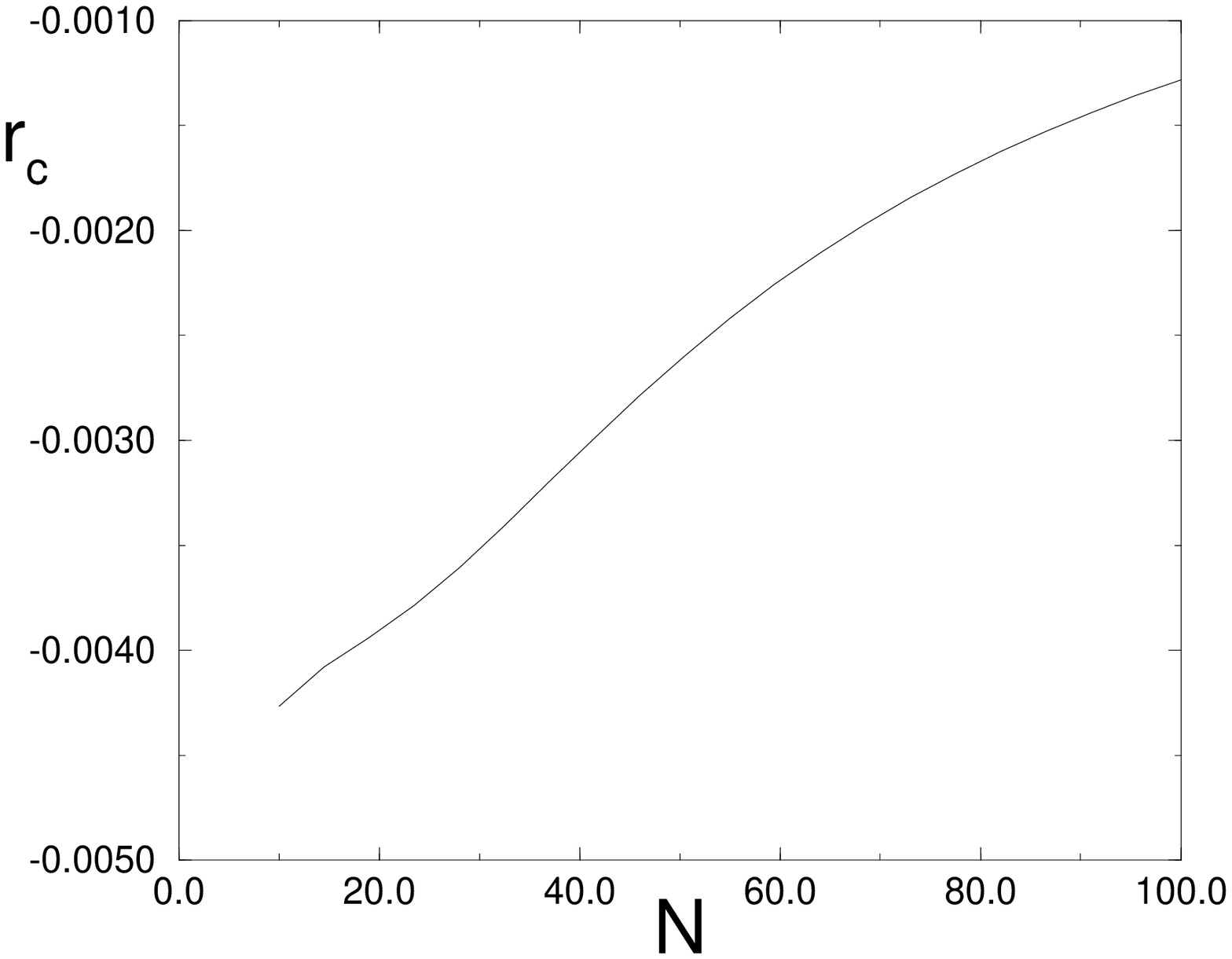} 
  \caption{\label{rc} Phase transition temperature $r_c(N)$ vs layer thickness 
  ($g_0=1 \cdot 10^{-5}, u_0=0.1$).}
\end{center}
\end{figure}

\section{Layer with out-of-plane easy axis}

As already mentioned, the propagator for the dipolar layer with out-of-plane 
easy axis,
\begin{eqnarray}
G_0(\bi{q},\bi{p})^{-1} =r_0 + c_0 \, k^2 - g (1+{\tilde{f}}_1) \, p -
g {\tilde{f}}_3 \, p^2 - g{\tilde{f}}_2 \, q^2 \, p \ , 
\end{eqnarray}
is not positive definite under the renormalisation group flow.
This indicates an instability of the homogeneous ground state at some finite
wave vector. 
Therefore this propagator cannot be used in a straightforward manner for the 
calculation of the critical behaviour of the system. 
Nevertheless we can draw some information regarding the instability from it. 
Thus, from the equation 
\begin{eqnarray}
\frac{\partial}{\partial p} \biggr|_{q=0} \left( r + c \, k^2 - 
g (1+\tilde{f}_1) \, p - g \tilde{f_3} \, p^2 - g \tilde{f_2} \, q^2 \, p
\right) = 0
\end{eqnarray}
we may infer the typical size of the structures in the low-temperature phase, 
as function of the dipole strength $g$ and the layer thickness $L$. 
The result is
\begin{eqnarray}
D_{\mbox{domain}} = \frac{2 \pi \left[ c + g \, \frac{L}{2} \left( 1 + 
\frac{L}{2a_0} \right) \right]}{g \left( 1 + \frac{L}{a_0} \right)} \ .
\label{domainsize} 
\end{eqnarray}
I.e., starting from a constant domain size at vanishing layer thickness, the
domain size decreases with increasing thickness, runs through a minimum, and
finally grows again linearly with $L$.
The result that a layer with out-of-plane easy axis should condense 
inhomogeneously is in accordance with Refs.~\cite{Kooy,Garel,Kaplan}; but the 
increase of the domain size is predicted to be proportional to the square root 
of the layer thickness in Refs.~\cite{Kooy,Kaplan}, and $\sim L^{1/3}$ in 
Ref.~\cite{Garel}.

\end{document}